# Nonlinear Multiphysics Modeling of Batch Digester Discharge Dynamics with Rheology-Driven Hydraulic Transport and Drainability Coupling

José M. Campos-Salazar
*Electronic Engineering Department*
*Universitat Politècnica de Catalunya*
Barcelona, Spain
jose.manuel.campos@upc.edu

ABSTRACT—**Batch digester blowdown operations constitute highly nonlinear hydraulic transport processes characterized by strong rheological coupling, evolving slurry consistency, nonlinear pressure–flow dynamics, and severe transport uncertainty. In this work, a comprehensive nonlinear dynamic model and robust sliding mode control (SMC) strategy are developed for the analysis and hydraulic regulation of industrial batch digester discharge systems operating under transient multiphase transport conditions. The proposed modeling framework incorporates coupled mass balances, consistency-dependent hydraulic resistance, non-Newtonian slurry rheology, mixture-density reconstruction, drainability dynamics, channeling effects, and nonlinear hydraulic transport behavior within a unified differential-algebraic formulation.**

**The nonlinear rheological behavior of the pulp suspension is represented through a Herschel–Bulkley constitutive formulation, while the hydraulic subsystem incorporates dynamic resistance reconstruction as a function of slurry consistency and transport conditions. To address the severe nonlinearities and uncertain operating conditions associated with industrial blowdown systems, a robust SMC architecture is proposed for discharge-flow regulation. The controller integrates an integral sliding manifold, nonlinear hydraulic actuation, boundary-layer regularization, supervisory protection mechanisms, and bounded hydraulic control laws in order to preserve robust tracking performance and stable closed-loop operation under strong rheological disturbances and transport uncertainties.**

**The complete nonlinear framework was implemented in MATLAB–Simulink using stiff numerical solvers and numerical regularization techniques specifically designed to avoid singularities, hydraulic runaway, and non-finite dynamic states during long-time transient simulations. The obtained results demonstrate that the proposed SMC strategy achieves robust discharge-flow regulation, rapid sliding-manifold convergence, strong disturbance rejection capability, and bounded hydraulic actuation despite severe variations in channeling behavior, drainability conditions, slurry consistency, and hydraulic transport resistance. Additionally, the energetic analysis reveals that the blowdown operation is dominated by strongly transient rheology-dependent hydraulic dissipation and consistency-driven transport losses. The proposed framework therefore establishes a physically consistent and computationally robust proof-of-concept methodology for advanced nonlinear modeling and robust hydraulic control of industrial batch digester discharge systems.**

## 1.0. INTRODUCTION

The industrial kraft pulping process constitutes one of the most complex thermochemical and transport-intensive operations within the pulp and paper industry due to the simultaneous interaction among multiphase flow dynamics, nonlinear rheology, heat transfer, chemical delignification reactions, liquor transport, and hydraulic discharge phenomena [1]–[4]. Among the various operational stages involved in kraft pulping, the batch digester discharge or blowdown operation represents a particularly critical transient process because it governs the rapid evacuation of concentrated pulp slurry and entrained black liquor from the digester vessel toward the blow tank through highly nonlinear hydraulic transport mechanisms [5]–[8]. During this operation, the internal fiber suspension experiences severe variations in consistency, permeability, rheological resistance, liquor redistribution, and hydraulic transport conditions, thereby generating strongly coupled nonlinear dynamics that substantially influence process stability, energetic consumption, equipment stress, discharge efficiency, and downstream washing performance [6], [9], [10].

Unlike conventional single-phase hydraulic transport systems, pulp-slurry blowdown processes involve concentrated non-Newtonian suspensions exhibiting yield-stress behavior, shear-thinning characteristics, dynamic permeability variation, and time-varying structural rearrangement of the fiber network [5], [11]–[13]. Experimental investigations reported in the literature have demonstrated that the rheological behavior of concentrated pulp suspensions depends strongly on consistency, fiber morphology, liquor retention, and suspension microstructure, leading to highly nonlinear pressure-drop characteristics and severe transport uncertainties during blow-line operation [11], [14], [15]. Furthermore, the formation of preferential flow paths, channeling phenomena, and dynamic drainage effects inside compacted fiber beds introduce additional nonlinearities and transport disturbances that cannot be adequately represented using conventional linearized hydraulic models [8], [12], [16].

Several studies have investigated dynamic modeling approaches for pulp transport systems, black-liquor circulation, dewatering operations, and batch digester hydraulics. Early modeling efforts were primarily focused on simplified mass balances, empirical hydraulic correlations, and steady-state transport formulations [3], [4], [17]. However, modern industrial requirements associated with process optimization, energy efficiency, advanced control, and digital process supervision have motivated the development of increasingly sophisticated nonlinear and multiphysics models capable of describing the coupled interaction among rheology, transport, consistency evolution, and hydraulic dynamics [6], [8], [10], [18]. Recent research has incorporated differential-algebraic formulations, non-Newtonian transport equations, consistency-dependent hydraulic resistance models, and distributed transport mechanisms in order to improve the predictive capability of pulp-processing simulations [8], [13], [17], [18].

Despite these advances, the majority of the available literature remains primarily focused on continuous digesters, bleaching systems, pulp drying, filtration systems, or simplified slurry transport pipelines [4], [18]–[21]. In contrast, relatively limited attention has been devoted to the transient nonlinear dynamics associated with batch digester discharge operations, particularly under severe rheological disturbances, evolving slurry consistency, and dynamically varying transport properties [6], [8], [10], [18]. Moreover, most existing hydraulic transport studies employ simplified flow assumptions, linear approximations, or steady-state formulations that neglect the strong nonlinear coupling between slurry composition, hydraulic resistance, and discharge-flow evolution [12], [18], [22]. Consequently, there remains a significant need for advanced nonlinear dynamic models capable of accurately representing the transient blowdown behavior of industrial batch digester systems under realistic operating conditions.

From a control-system perspective, the regulation of batch digester discharge processes constitutes a particularly challenging problem due to the presence of strong nonlinearities, time-varying parameters, multiphase interactions, uncertain rheological properties, nonlinear actuator dynamics, and external disturbances associated with channeling and drainability variations [7], [10], [23]. Conventional linear control approaches, including proportional–integral–derivative (PID) regulators and classical feedback controllers, often exhibit degraded performance when applied to strongly nonlinear hydraulic systems operating under large parametric uncertainty and severe transport disturbances [24]–[26]. In particular, the effective hydraulic resistance of concentrated pulp suspensions may vary several orders of magnitude during discharge due to evolving consistency and structural changes in the fiber network, thereby producing substantial variations in the underlying process dynamics [11], [14], [15].

For highly nonlinear industrial processes, robust nonlinear control methodologies have increasingly attracted attention due to their superior disturbance-rejection capability and robustness against modeling uncertainty [26]–[28]. Among these techniques, sliding mode control (SMC) has emerged as one of the most effective robust nonlinear control strategies for uncertain dynamic systems characterized by strong nonlinearities, parameter variations, matched disturbances, and uncertain transport dynamics [29], [29], [30]. The principal advantage of SMC lies in its capability to enforce system trajectories onto a predefined sliding manifold, thereby producing closed-loop dynamics that remain largely insensitive to bounded uncertainties and external disturbances [29], [29], [30].

The SMC has been successfully applied to numerous nonlinear engineering systems, including robotic manipulators, chemical reactors, electric drives, power electronic converters, hydraulic actuators, multiphase flow systems, and nonlinear transport processes [24], [26], [28]–[30]. In hydraulic and process-engineering applications, SMC has demonstrated excellent robustness under severe parameter uncertainty and nonlinear flow conditions, particularly in systems exhibiting nonlinear pressure–flow coupling, uncertain transport coefficients, and actuator nonlinearities [24], [26], [28]–[30]. Additionally, modern boundary-layer and continuous approximation formulations have substantially mitigated the classical chattering problem traditionally associated with discontinuous switching control laws [24], [29], [31].

Nevertheless, despite the extensive literature available on SMC theory and nonlinear hydraulic control, very limited research has investigated the application of robust sliding-mode strategies to nonlinear batch digester discharge systems involving concentrated pulp-slurry transport, consistency-dependent rheology, dynamic drainage phenomena, and nonlinear hydraulic blowdown behavior [6], [8], [32]. Furthermore, the coupled interaction among channeling effects, drainability disturbances, evolving slurry density, and nonlinear transport resistance has not been comprehensively integrated into an SMC-oriented nonlinear dynamic modeling framework for industrial pulp-processing applications.

Motivated by these limitations, the present work proposes a comprehensive nonlinear dynamic model and robust SMC architecture for the hydraulic regulation of batch digester discharge systems operating under highly nonlinear rheological transport conditions. The proposed model incorporates coupled mass balances, consistency-dependent hydraulic resistance, non-Newtonian transport behavior, dynamic slurry-density reconstruction, channeling effects, drainability disturbances, and nonlinear hydraulic actuator dynamics within a unified differential-algebraic framework. Subsequently, a robust SMC strategy is developed in order to regulate the discharge flow while preserving stability and disturbance rejection under severe nonlinear operating conditions.

It is important to emphasize that the present investigation is conceived as a proof-of-concept study focused on the development, numerical implementation, and theoretical evaluation of a nonlinear dynamic modeling and robust control framework for batch digester discharge systems. Consequently, the primary objective of this work is not the experimental validation of a specific industrial installation, but rather the demonstration of the feasibility, robustness, numerical stability, and control applicability of the proposed nonlinear hydraulic transport formulation and SMC architecture under representative operating scenarios and severe rheological disturbances. The adopted simulation-oriented methodology enables the systematic evaluation of the coupled nonlinear interactions among slurry consistency, hydraulic resistance, drainability effects, channeling phenomena, and discharge-flow regulation, thereby establishing a theoretical and computational foundation for future industrial-scale implementation, parameter identification, and experimental validation studies.

The principal contributions of this work can therefore be summarized as follows:

1. Development of a comprehensive nonlinear dynamic model for batch digester blowdown operation incorporating multiphase transport, nonlinear rheology, dynamic drainage behavior, and consistency-dependent hydraulic resistance.

2. Integration of channeling effects and drainability disturbances into the nonlinear hydraulic transport formulation.

3. Formulation of a robust SMC architecture specifically designed for nonlinear discharge-flow regulation under severe rheological uncertainty and transport disturbances.

4. Implementation of numerical regularization and protection mechanisms enabling stable long-time simulation of the strongly nonlinear differential-algebraic process.

5. Evaluation of the proposed controller through transient simulations under severe nonlinear disturbances and evolving transport conditions.

The obtained results demonstrate that the proposed nonlinear SMC framework provides robust discharge-flow tracking, bounded hydraulic actuation, strong disturbance rejection capability, and stable energetic behavior despite the highly nonlinear and strongly coupled nature of industrial batch digester discharge systems.

The remainder of this manuscript is organized as follows. Section 2 presents the process description of the nonlinear batch digester discharge system. Section 3 develops the nonlinear multiphysics modeling framework, while Section 4 introduces the thermodynamic and hydraulic efficiency formulation. Section 5 presents the proposed SMC architecture and stability analysis. Section 6 discusses the transient simulation results under nonlinear rheological disturbances. Finally, Section 7 summarizes the principal conclusions and future research directions.

## 2.0. PROCESS DESCRIPTION

The batch digester discharge system illustrated in Fig. 1 constitutes a highly nonlinear, multivariable, and strongly coupled transport process in which hydraulic flow behavior, fiber-network dynamics, liquor redistribution, and rheological evolution occur simultaneously and interact through state-dependent feedback mechanisms [1], [33], [34]. From a process-systems perspective, the discharge operation represents a transient multiphase hydrotransport phenomenon involving the continuous evolution of solid and liquid inventories under severe rheological and hydraulic constraints. The overall system boundary comprises the batch digester vessel, the discharge extraction region, the hydraulic transport pipeline, the pump subsystem, and the downstream blow-tank connection, all of which participate dynamically in the global discharge behavior.

Within the proposed framework, the digester is represented as a lumped nonlinear control volume operating under transient conditions, where the internal process dynamics are governed primarily by the temporal evolution of the dry-fiber mass $M_s(t)$ and the free-liquor mass $M_{fl}(t)$. These state variables establish the instantaneous thermodynamic, hydraulic, and rheological condition of the pulp suspension inside the vessel and determine the evolution of secondary variables such as pulp consistency $C(t)$, phase distribution, mixture density $\rho_{mix}(t)$, liquid holdup, fiber occupancy, and total slurry volume $V(t)$. The dynamic interaction among these variables generates a highly coupled transport environment in which changes in one physical domain immediately propagate through the remaining subsystems. Such coupled dynamic behavior is characteristic of industrial kraft pulping operations and has been extensively identified in modern digester modeling methodologies for both continuous and batch configurations [1], [33]–[35].

As shown in Fig. 1, the digester receives an inlet dilution-liquor stream $f_{in}(t)$, which directly modifies the internal liquid inventory and indirectly alters the structural and rheological properties of the pulp suspension. The dilution stream therefore plays a dual physical role: first, as a hydraulic inventory input influencing the total liquid accumulation within the vessel, and second, as a rheological-conditioning mechanism affecting slurry consistency, fiber-network permeability, apparent viscosity, and effective hydraulic transport resistance. The assumption of homogeneous mixing and lumped dynamic representation adopted in the proposed model is fully consistent with advanced control-oriented representations commonly employed in pulp and paper process modeling, where spatially distributed phenomena are transformed into equivalent nonlinear dynamic states suitable for transient simulation and control-system development [1], [33]–[35].

Additionally, the inlet dilution-liquor flow $f_{in}(t)$ plays a critical role in the transient hydraulic behavior of the discharge process because it continuously modifies the internal liquid inventory, slurry consistency, and effective rheological transport conditions inside the digester. Variations in $f_{in}(t)$ directly affect suspension fluidization, apparent viscosity, hydraulic resistance, and transport capability, thereby influencing the discharge-flow dynamics throughout the blowdown operation. Experimental studies on pulp hydrotransport systems have demonstrated that dilution conditions strongly affect pressure-drop behavior, slurry mobility, and transport efficiency in concentrated fibers suspensions [36]–[40].

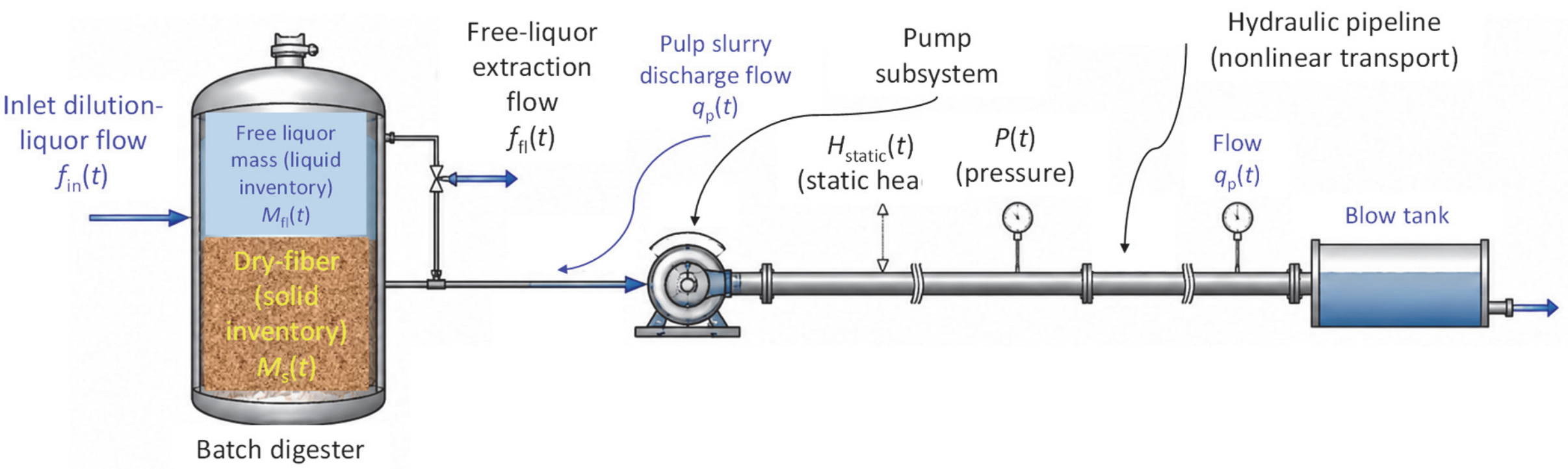

Fig. 1. Physical representation of the nonlinear batch digester discharge system, including the dilution-liquor inlet, free-liquor extraction, pulp-slurry transport line, pump subsystem, and blow tank hydraulic discharge configuration.

Two physically distinct outlet streams emerge from the digester during the discharge process. The first corresponds to the free-liquor extraction stream $f_{fl}(t)$, associated with liquid drainage and liquor release from the fiber matrix, while the second corresponds to the pulp slurry discharge flow $q_p(t)$, which transports the concentrated pulp suspension through the hydraulic pipeline toward the blow tank. Although these outlet mechanisms are physically different, they remain strongly interconnected through the internal consistency dynamics and the evolving structure of the fiber network. Consequently, the discharge system cannot be interpreted as a set of independent flows, but rather as a fully coupled nonlinear transport process governed by the interaction between hydraulic forcing, suspension rheology, and internal liquor redistribution.

The pulp consistency $C(t)$ acts as the principal coupling variable linking the mass-balance subsystem with the nonlinear hydraulic transport dynamics. Variations in consistency directly affect the effective rheological properties of the suspension, including apparent viscosity, yield stress, plug-flow behavior, and shear-thinning characteristics, thereby altering the hydraulic resistance experienced by the discharge flow. As the discharge progresses and the internal liquid inventory decreases, the pulp suspension gradually transitions toward increasingly concentrated operating conditions, producing substantial modifications in transportability and flow resistance. This phenomenon introduces severe nonlinearities into the hydraulic subsystem and establishes a dynamic feedback mechanism between inventory depletion and discharge capability.

The rheological behavior of the pulp suspension is represented according to non-Newtonian constitutive frameworks commonly employed for concentrated fiber suspensions, particularly Herschel–Bulkley-type formulations capable of describing simultaneous yield-stress and shear-thinning effects [36]–[39]. Such rheological behavior is characteristic of industrial kraft pulp suspensions and has been experimentally verified in numerous studies addressing pulp-fiber hydrotransport, pressure-drop dynamics, suspension mobility, and pipeline transport phenomena [36]–[38], [40]. Experimental investigations reported in the literature demonstrate that the apparent viscosity and effective hydraulic resistance of pulp suspensions exhibit strong dependence on consistency level, fiber concentration, shear conditions, suspension morphology, and flow regime [36]–[38], [40]. Consequently, relatively small variations in pulp consistency may produce substantial changes in the hydraulic response of the discharge system, particularly during high-consistency blowdown operations.

The hydraulic subsystem shown in Fig. 1 consists of the pump and pipeline arrangement responsible for transforming the applied hydraulic head into slurry transport flow. This subsystem establishes a nonlinear hydraulic coupling between the pump actuation and the discharge flow through a consistency-dependent pressure-flow relationship. Due to the nonlinear rheological characteristics of the pulp suspension, the resulting hydraulic transport mechanism cannot be represented through conventional linear flow equations. Instead, the discharge dynamics exhibit strongly nonlinear behavior associated with variable apparent viscosity, yield-stress effects, evolving friction losses, and state-dependent hydraulic resistance. Similar nonlinear hydraulic couplings have been extensively discussed in advanced process-oriented and control-oriented pulp transport models reported in the literature [35], [39], [41], [42].

Additionally, Fig. 1 highlights the incorporation of phenomenological transport effects that further influence the discharge dynamics, particularly the channeling factor $k_{ch}(t)$ and the drainability coefficient $y_K(t)$. The channeling factor represents the formation of preferential flow pathways through the fiber matrix, allowing portions of the liquor phase to bypass the bulk suspension structure. Physically, this phenomenon is associated with local heterogeneities and structural rearrangements within the fiber network during displacement and dewatering operations. The existence of channeling significantly modifies the effective liquid transport pathways inside the digester and introduces additional nonlinearities into the overall discharge process.

Similarly, the drainability coefficient characterizes the time-dependent capability of the fiber matrix to release entrained liquor during hydraulic transport and displacement operations. This parameter reflects the evolving permeability and compressibility behavior of the pulp network and captures the progressive changes in fiber-network structure occurring throughout the blowdown process. Experimental and theoretical studies focused on pulp permeability, drainage phenomena, and suspension dewatering have demonstrated that fiber-network restructuring plays a dominant role in determining liquid-retention characteristics and hydraulic transport behavior during high-consistency operations [36], [40], [43]. Consequently, the incorporation of channeling and drainability effects enables the proposed model to reproduce important industrial phenomena frequently neglected in simplified discharge representations.

The combined interaction between dry-fiber inventory, free-liquor inventory, pulp consistency, rheological resistance, hydraulic transport, drainability behavior, and channeling effects produces a strongly interconnected nonlinear dynamic system exhibiting differential-algebraic characteristics. Variations in $M_s(t)$ and $M_{fl}(t)$ propagate through the consistency dynamics and subsequently modify the hydraulic transport conditions, which in turn influence the discharge flow and feed back into the original mass balances. This recursive interaction establishes a closed-loop physical coupling between the hydraulic and process subsystems, consistent with the behavior observed in advanced nonlinear digester discharge models and complex industrial slurry transport systems [35], [39], [41], [42].

Therefore, the proposed process description provides a physically consistent and control-oriented representation of the batch digester discharge operation, capable of capturing the dominant nonlinear interactions governing slurry transport, hydraulic behavior, rheological evolution, and liquor redistribution under industrial operating conditions.

Finally, the remainder of this manuscript is organized as follows. Section 2 presents the physical process description of the nonlinear batch digester discharge system, including the principal hydraulic, rheological, and transport mechanisms governing the blowdown operation. Section 3 develops the comprehensive nonlinear multiphysics modeling framework based on coupled mass balances, consistency-dependent hydraulic transport, rheological resistance reconstruction, and phenomenological drainability and channeling effects. Subsequently, Section 4 introduces the thermodynamic and hydraulic efficiency formulation associated with the nonlinear discharge process. Section 5 presents the proposed robust SMC architecture, including the nonlinear control-oriented formulation, sliding-surface synthesis, Lyapunov stability analysis, robustness properties, chattering mitigation strategy, and supervisory protection mechanisms. Section 6 discusses the transient simulation results obtained under severe nonlinear rheological disturbances and evolving transport conditions, including the hydraulic, energetic, and robustness analyses of the proposed control framework. Finally, the principal conclusions and future research directions are presented in the last section of the manuscript.

## 3.0. SYSTEM MODELING

The batch digester discharge system illustrated in Fig. 1 is modeled as a nonlinear, multiphysics, and strongly coupled dynamic process governed by the simultaneous interaction of mass conservation, rheological transport, hydraulic resistance, and phenomenological drainage effects. The proposed formulation is based on a lumped-parameter representation in which the digester is treated as a perfectly mixed nonlinear control volume, while the downstream blow-line transport subsystem is represented through a nonlinear hydraulic relation coupling slurry consistency, rheological evolution, and hydraulic discharge behavior. Similar nonlinear formulations have been widely employed in advanced kraft digester and pulp hydrotransport models [1], [33]–[35], [38], [39], [41]–[43].

The resulting mathematical structure corresponds to a nonlinear differential-algebraic system because the slurry discharge flow depends implicitly on the evolving rheological and hydraulic conditions of the suspension. The dynamic interaction between inventory depletion, consistency evolution, and hydraulic transport establishes a recursive feedback mechanism characteristic of highly concentrated pulp-suspension systems operating under transient blowdown conditions.

The instantaneous slurry consistency is reconstructed according to

$$C(t) = M_s(t)/(M_s(t) + M_{fl}(t) + \varepsilon) \tag{1}$$

where $\varepsilon$ represents a numerical regularization parameter introduced to avoid singularities during simulation and nonlinear hydraulic reconstruction. The consistency variable constitutes the principal nonlinear coupling mechanism linking the inventory subsystem with the rheological and hydraulic transport dynamics.

The total slurry mass inside the digester is reconstructed as

$$M_{\mathrm{total}}(t) = M_{\mathrm{s}}(t) + M_{\mathrm{fl}}(t) \tag{2}$$

while the total occupied volume is obtained from the phase contributions according to

$$V(t) = V_{\mathrm{s}}(t) + V_{\mathrm{fl}}(t) \tag{3}$$

where the liquid and solid phase volumes are represented by

$$V_{\mathrm{fl}}(t) = M_{\mathrm{fl}}(t)/\rho_{\mathrm{fl}} \tag{4}$$

and

$$V_{\mathrm{s}}(t) = M_{\mathrm{s}}(t)/(\rho_{\mathrm{s}}\cdot(1 - w)) \tag{5}$$

respectively. Similar phase-reconstruction methodologies are extensively employed in nonlinear pulp-suspension and digester transport models because they allow dynamic estimation of slurry occupancy, phase redistribution, and hydraulic loading conditions [1], [33], [34], [43].

The effective slurry density is reconstructed from the phase distributions as

$$\rho_{\mathrm{mix}}(t) = \frac{M_{\mathrm{s}}(t) + M_{\mathrm{fl}}(t)}{\frac{M_{\mathrm{s}}(t)}{\rho_{\mathrm{s}}} + \frac{M_{\mathrm{fl}}(t)}{\rho_{\mathrm{fl}}} + \varepsilon} \tag{6}$$

which introduces additional nonlinear coupling between the inventory dynamics and the hydraulic subsystem. Because the effective mixture density evolves continuously throughout the blowdown process, the corresponding hydraulic transport conditions become strongly time varying.

The nonlinear dry-fiber inventory dynamics are represented through

$$\frac{\mathrm{d}M_{\mathrm{s}}(t)}{\mathrm{d}t} = -f_{\mathrm{s}}(t) \tag{7}$$

where the discharged fiber transport flow is modeled as

$$f_{\mathrm{s}}(t) = \rho_{\mathrm{mix}}(t)\cdot C(t)\cdot q_{\mathrm{p}}(t) \tag{8}$$

thereby coupling the dry-fiber transport directly with the hydraulic discharge flow and the evolving slurry consistency. This formulation assumes homogeneous suspension transport within the blow-line subsystem and reproduces the dominant transport behavior observed in industrial slurry discharge systems [35], [38], [41].

Similarly, the nonlinear free-liquor inventory balance is formulated as

$$\frac{\mathrm{d}M_{\mathrm{fl}}(t)}{\mathrm{d}t} = \rho_{\mathrm{fl}}\cdot f_{\mathrm{in}}(t) - \rho_{\mathrm{fl}}\cdot f_{\mathrm{fl}}(t) - f_{\mathrm{liq}}(t) \tag{9}$$

where the free-liquor transport associated with the discharged pulp suspension is represented through

$$f_{\mathrm{liq}}(t) = (1 - k_{\mathrm{ch}}(t))\cdot(1 - y_{\mathrm{k}}(t)\cdot C(t))\cdot\rho_{\mathrm{mix}}(t)\cdot(1 - C(t))\cdot q_{\mathrm{p}}(t) \tag{10}$$

Equation (10) incorporates phenomenological corrections associated with preferential flow paths, dynamic liquor-retention mechanisms, and transient permeability effects observed in concentrated pulp suspensions [7], [36], [39], [43]–[46]. The channeling factor introduces the effects of preferential hydraulic pathways through the fiber matrix, while the drainability coefficient represents the time-dependent capability of the fiber network to release entrained liquor during displacement and blowdown operations.

The rheological behavior of the pulp suspension is represented through the Herschel–Bulkley constitutive framework,

$$\tau(t) = \tau_{\mathrm{y}} + K_{\mathrm{HB}}\cdot\dot{\gamma}^{n}(t) \tag{11}$$

which reproduces the combined yield-stress and shear-thinning behavior characteristic of concentrated fiber suspensions. Experimental investigations on pulp hydrotransport and non-Newtonian slurry systems have demonstrated that the apparent viscosity and pressure-drop characteristics depend strongly on slurry consistency, fiber concentration, and flow regime [36]–[40]. Consequently, the hydraulic transport subsystem becomes directly coupled with the evolving rheological condition of the pulp suspension.

The consistency-dependent hydraulic resistance is represented as

$$C_n(t) = K_{\mathrm{ref}} \cdot \left( \frac{C(t) + \varepsilon}{C_{\mathrm{ref}}} \right)^{\alpha_C} \tag{12}$$

where the nonlinear exponent $\alpha_C$ determines the sensitivity of the hydraulic resistance to consistency variations. This formulation reflects the strong nonlinear increase in transport resistance typically observed in concentrated pulp suspensions and highly loaded slurry pipelines [7], [36]–[40], [44]–[47].

The static hydraulic head generated by the slurry column is represented through

$$H_{\mathrm{static}}(t) = K_{\mathrm{static}} \cdot \rho_{\mathrm{mix}}(t) \tag{13}$$

thereby introducing direct coupling between mixture-density evolution and hydraulic loading conditions.

The nonlinear hydraulic transport subsystem is represented through a quasi-steady pressure-flow relation coupling the applied hydraulic actuation head with the discharge flow,

$$q_{\mathrm{p}}^{\mathrm{alg}}(t) = \left( \frac{\max(H_0(t) - H_{\mathrm{static}}(t), 0)}{C_n(t) + \varepsilon} \right)^{\frac{1}{n}} \tag{14}$$

which reveals the strong nonlinear interaction between rheology, hydraulic resistance, and slurry transport behavior. Similar nonlinear pump-flow couplings have been extensively reported in pulp hydrotransport systems and non-Newtonian multiphase transport applications [7], [37], [38], [40], [41], [44]–[47].

To eliminate algebraic loops and improve numerical robustness during long-time nonlinear simulations, the hydraulic transport subsystem is regularized through a first-order hydraulic relaxation model,

$$\frac{\mathrm{d}q_{\mathrm{p}}(t)}{\mathrm{d}t} = \frac{q_{\mathrm{p}}^{\mathrm{alg}}(t) - q_{\mathrm{p}}(t)}{\tau_{\mathrm{p}}} \tag{15}$$

where $\tau_{\mathrm{p}}$ represents the effective hydraulic relaxation time associated with pump inertia, transport delay, and hydraulic transient dynamics. The inclusion of this relaxation mechanism significantly improves numerical stability in the presence of severe rheological nonlinearities and rapidly evolving transport conditions.

The resulting formulation therefore establishes a strongly coupled nonlinear dynamic system in which variations in the internal inventories continuously modify the slurry consistency, which subsequently alters the rheological resistance, effective mixture density, static hydraulic head, and discharge-flow dynamics. The updated hydraulic transport conditions then feed back into the inventory balances, producing a recursive nonlinear interaction characteristic of industrial batch digester blow-line operations and complex pulp transport systems [1], [33]–[35], [41]–[43].

Consequently, the proposed modeling framework provides a physically consistent and control-oriented representation of the batch digester discharge operation capable of reproducing the dominant nonlinear interactions governing rheological evolution, hydraulic transport, liquor redistribution, permeability variation, and transient slurry discharge behavior under realistic industrial operating conditions.

### 4.0. THERMODYNAMIC AND HYDRAULIC EFFICIENCY FORMULATION

The energetic and hydraulic performance of the batch digester discharge system constitutes one of the most critical aspects governing transport capability, operational stability, hydraulic loading, and overall process efficiency during blow-line operation in industrial kraft pulp production systems. In highly concentrated pulp-discharge operations, the global system efficiency cannot be interpreted exclusively from a conventional pump-efficiency perspective because the discharge dynamics are strongly influenced by multiphase transport interactions, non-Newtonian rheology, consistency-dependent hydraulic resistance, liquor-retention phenomena, and dynamic restructuring of the fiber network [36]–[44]. Consequently, the effective energetic performance emerges from the coupled interaction between hydraulic energy conversion, rheological transport behavior, slurry mobilization capability, and the dynamic release of entrained liquor during discharge operations.

From a thermodynamic and hydraulic standpoint, the total hydraulic power transferred to the pulp suspension is governed by the interaction between the applied hydraulic head and the evolving discharge flow. Accordingly, the instantaneous hydraulic transport power is represented as

$$P_{\mathrm{h}}(t) = H_0(t)\cdot q_{\mathrm{p}}(t) \tag{16}$$

where the hydraulic power evolves dynamically according to the nonlinear consistency-dependent hydraulic subsystem described previously. Since both the hydraulic actuation and the discharge flow are strongly coupled with the evolving rheological condition of the suspension, the energetic behavior of the system becomes intrinsically nonlinear, time varying, and state dependent [37], [38], [40]–[42]. Similar hydraulic-power formulations have been extensively employed in pulp hydrotransport systems, slurry pipeline networks, and concentrated multiphase transport analyses involving non-Newtonian fiber suspensions [36]–[40].

The effective hydraulic transport efficiency may therefore be interpreted as the ratio between the useful transport power associated with effective slurry conveyance and the total hydraulic power supplied by the hydraulic subsystem. The instantaneous transport efficiency is consequently defined as

$$\eta_{\mathrm{h}}(t) = P_{\mathrm{useful}}(t)/P_{\mathrm{h}}(t) \tag{17}$$

where the useful energetic contribution is associated with the effective mobilization and transport of both the dry-fiber phase and the entrained liquor phase toward the blow tank. Conversely, hydraulic losses arise from viscous dissipation, rheology-driven pressure losses, turbulent transport effects, channeling phenomena, internal suspension friction, fiber-network deformation, and dynamic permeability variations occurring throughout the discharge operation [36]–[40], [44].

Since the pulp suspension exhibits highly concentrated non-Newtonian behavior characterized by Herschel–Bulkley rheology, the effective transport efficiency becomes strongly dependent on the evolving consistency conditions inside the digester. Experimental investigations on concentrated pulp hydrotransport demonstrate that increases in consistency produce substantial increases in apparent viscosity, yield stress, and pressure-drop requirements, thereby reducing the effective hydraulic efficiency of the transport process [36], [37], [39], [40]. This effect becomes particularly severe during batch digester blowdown operations because the progressive depletion of free liquor continuously transforms the suspension into an increasingly compact and hydraulically resistant fiber network requiring progressively larger hydraulic energy for transport.

The rheological dissipation associated with slurry transport is represented through the viscous dissipation functional

$$\Phi_{\mathrm{v}}(t) = \tau(t)\cdot\dot{\gamma}(t) \tag{18}$$

which quantifies the instantaneous volumetric dissipation rate associated with internal shear deformation and rheological energy losses. This dissipation mechanism constitutes one of the dominant energetic loss contributions in high-consistency pulp transport systems and has been extensively investigated in recent studies involving fibrous suspensions, non-Newtonian slurry transport, and concentrated hydrotransport systems [7], [36]–[38], [40], [44]–[46]. As the suspension consistency increases, the corresponding shear stresses and internal friction effects intensify significantly, producing a substantial increase in hydraulic energy consumption.

The energetic behavior of the system is additionally influenced by the dynamic drainability coefficient, which introduces a direct coupling between fiber-network permeability and hydraulic energy requirements. As the drainability capability decreases, larger quantities of entrained liquor remain trapped within the fiber structure, thereby increasing the effective apparent viscosity and hydraulic transport resistance of the suspension. Consequently, the discharge system requires progressively larger hydraulic power to sustain equivalent discharge-flow conditions under degraded drainage behavior. Similar coupling mechanisms between permeability reduction, drainage degradation, and elevated energetic consumption have been experimentally reported in pulp dewatering systems and fiber-network transport studies [7], [40], [43]–[46].

The channeling phenomenon also exerts a significant influence on the energetic performance of the discharge operation. Physically, channeling produces preferential low-resistance pathways through the fiber matrix, allowing portions of the liquor phase to bypass the bulk suspension structure. Although such pathways may locally reduce hydraulic pressure losses, they simultaneously deteriorate transport uniformity and reduce the effectiveness of momentum transfer between the liquid and solid phases. Consequently, the global transport efficiency becomes governed not only by hydraulic energy consumption, but also by the effectiveness of fiber mobilization, suspension homogenization, and momentum distribution throughout the discharge system [43], [45], [46].

The nonlinear dependence between hydraulic resistance and slurry consistency further intensifies the energetic coupling within the system. The effective hydraulic resistance evolves according to (12), which demonstrates that the transport resistance increases nonlinearly as the suspension becomes more concentrated. As a consequence, the discharge flow progressively decreases during blowdown operation because increasingly larger hydraulic energy is required to overcome the elevated rheological resistance associated with high-consistency transport conditions.

Similar nonlinear efficiency degradation mechanisms have been widely reported in advanced slurry pipeline systems, industrial hydrotransport operations, and multiphase pulp transport networks [7], [37], [38], [40]–[42], [45]–[47].

The consistency-dependent hydraulic transport relation governing the discharge subsystem is represented in (14), which reveals that the energetic behavior of the process depends simultaneously on hydraulic forcing, rheological resistance, static hydraulic loading, and consistency evolution. Under elevated consistency conditions, the nonlinear increase in $C_n(t)$ produces substantial reductions in discharge capability, thereby increasing the hydraulic energy required to sustain equivalent transport rates.

Additionally, the hydraulic transport subsystem is dynamically regularized according to (15), thereby introducing transient hydraulic inertia effects associated with pump dynamics and blow-line transport propagation. These transient effects further contribute to the energetic response of the system because hydraulic acceleration and flow adaptation mechanisms generate additional dynamic power requirements during severe transport transients.

The combined interaction between rheological dissipation, consistency-dependent hydraulic resistance, channeling behavior, permeability degradation, and hydraulic transport dynamics therefore establishes a highly nonlinear energetic environment in which the effective efficiency continuously evolves throughout the blowdown operation. Consequently, the global energetic behavior of the discharge process cannot be represented through conventional steady-state efficiency formulations, but instead requires a fully coupled nonlinear transport representation capable of reproducing the dynamic interaction between hydraulic energy conversion, suspension rheology, and multiphase transport phenomena under industrial operating conditions [7], [36]–[47].

## 5.0. CONTROL SYSTEM DESIGN

The nonlinear batch digester discharge system described previously exhibits strong nonlinearities, rheology-dependent hydraulic coupling, parametric uncertainty, and time-varying transport dynamics arising from the interaction among slurry consistency evolution, hydraulic resistance variation, liquor-retention phenomena, and nonlinear hydrotransport behavior. The dynamic interaction among rheological transport mechanisms, inventory depletion, and pump-driven discharge flow generates a highly coupled nonlinear process whose operating conditions evolve continuously throughout the blowdown phase. Consequently, conventional linear control methodologies become insufficient to guarantee robust discharge regulation under severe operating variations and nonlinear rheological disturbances. Similar control limitations have been extensively reported in nonlinear slurry transport systems, non-Newtonian hydraulic networks, and multiphase industrial transport processes [24], [25], [27], [30], [31], [48].

To overcome these limitations, a SMC strategy is proposed for regulating the discharge flow during batch digester blowdown operation. Sliding-mode methodologies are particularly attractive for nonlinear industrial systems because they provide strong robustness against parametric uncertainty, unmatched disturbances, nonlinear coupling, model inaccuracies, and time-varying operating conditions [24], [25], [27], [30], [31], [48]. In hydraulic and slurry-transport applications, SMC structures have demonstrated superior robustness compared with conventional PI and PID control strategies, particularly under severe rheological variations and uncertain transport coefficients [24], [25], [27], [30], [31], [48].

As illustrated in Fig. 2, the proposed SMC framework regulates exclusively the discharge flow, while the remaining process variables evolve naturally according to the internal nonlinear process dynamics. This control philosophy is particularly suitable for batch digester blowdown operations because the discharge flow constitutes the dominant manipulated variable governing hydraulic transport capability, suspension mobilization, rheological transport behavior, and the overall discharge progression of the pulp slurry. Unlike conventional multivariable approaches attempting to regulate all internal inventories simultaneously, the proposed methodology intentionally focuses on direct hydraulic discharge regulation, thereby preserving the intrinsic physical coupling among slurry consistency, effective mixture density, hydraulic resistance, and fiber-network evolution.

The complete control architecture is organized into four principal subsystems: (A) supervisory inventory and consistency management layer, (B) nonlinear hydraulic SMC subsystem, (C) nonlinear batch digester discharge process model, and (D) monitored process outputs and diagnostic variables. The supervisory layer generates a protected discharge-flow reference based on process-operability constraints associated with hydraulic transportability and suspension consistency limitations. The nonlinear hydraulic SMC subsystem then computes the hydraulic actuation required to maintain robust discharge-flow tracking under severe nonlinear disturbances and rheological uncertainty.

*5.1. Nonlinear control-oriented representation*

The dominant hydraulic transport dynamics are represented through the nonlinear hydraulic relaxation subsystem introduced previously, together with the consistency-dependent hydraulic reconstruction and the nonlinear pressure-flow coupling governing the blow-line transport behavior. The resulting hydraulic subsystem becomes nonlinear, time varying, and strongly coupled with the evolving internal process states because the effective hydraulic transport gain depends continuously on the dynamic evolution of slurry consistency, rheological resistance, channeling behavior, and drainability conditions.

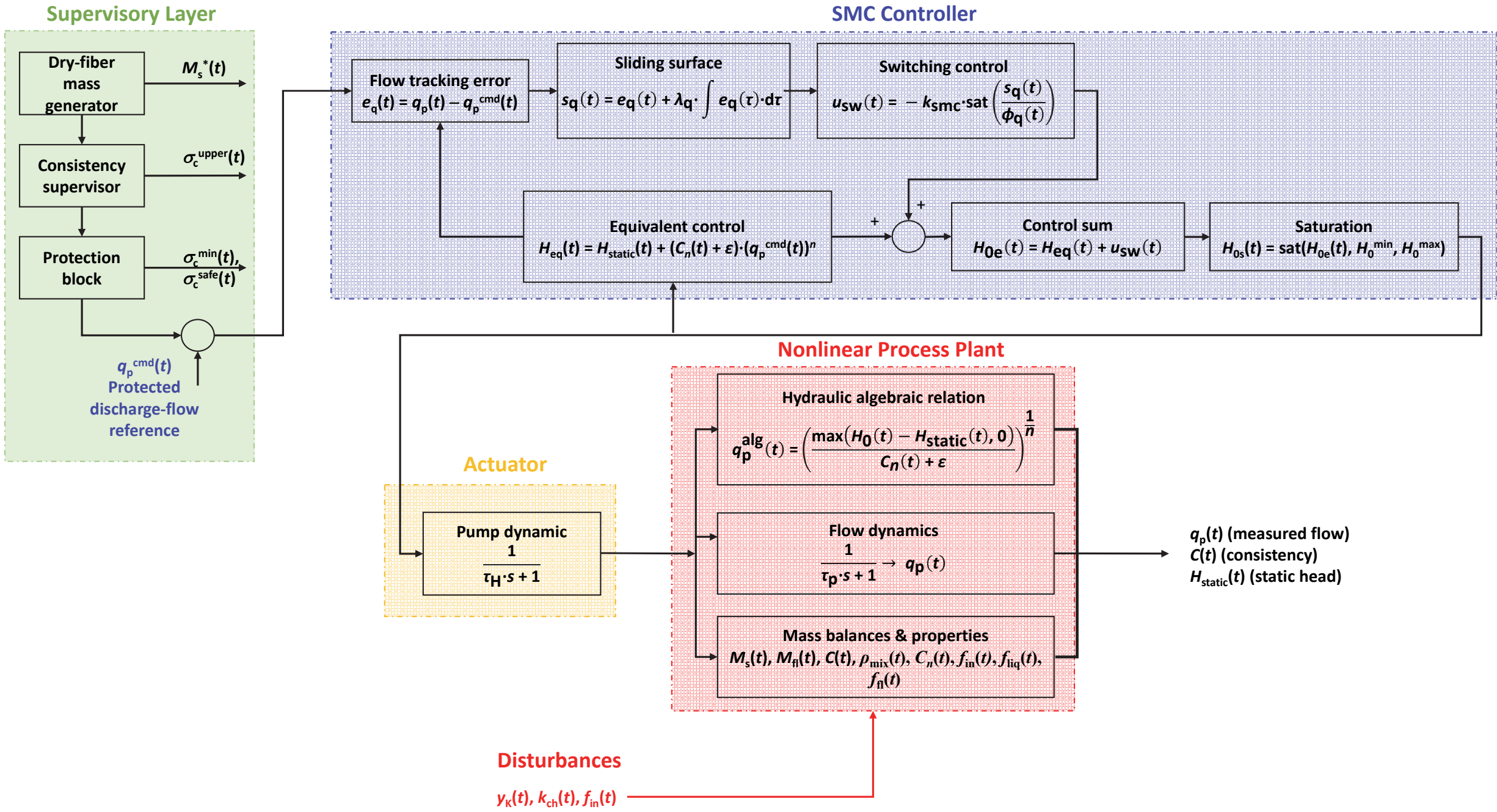

Fig. 2. Proposed SMC architecture for nonlinear discharge-flow regulation in the batch digester system, including the supervisory layer, hydraulic actuator dynamics, nonlinear process model, and disturbance-rejection structure.

The consistency reconstruction is expressed in (1), while the nonlinear hydraulic transport dynamics become

$$\frac{\mathrm{d}q_{\mathrm{p}}(t)}{\mathrm{d}t}=\frac{1}{\tau_{\mathrm{p}}}\cdot\left[\left(\frac{H_0(t)-H_{\mathrm{static}}(t)}{C_n(t)+\varepsilon}\right)^{\frac{1}{n}}-q_{\mathrm{p}}(t)\right] \tag{19}$$

Equations (12), (15), and (19) demonstrate that the effective hydraulic gain evolves continuously throughout the discharge operation because the internal slurry structure and rheological conditions remain strongly time dependent.

The nonlinear transport dynamics are additionally affected by channeling disturbances, drainability degradation, and hydraulic inventory variations, all of which continuously modify the effective discharge capability of the suspension. Consequently, the control problem becomes particularly challenging because the hydraulic subsystem operates under uncertain and continuously evolving transport conditions.

*5.2. Control Objective*

The primary control objective consists of forcing the discharge flow to track a desired supervisory reference despite the presence of:

- nonlinear rheological effects,
- consistency-dependent hydraulic resistance,
- uncertain slurry properties,
- channeling disturbances,
- drainability variations,
- hydraulic nonlinearities,
- actuator saturation,
- and model uncertainty.

Accordingly, the flow-tracking error is defined as

$$e_{\mathrm{q}}(t) = q_{\mathrm{p}}(t) - q_{\mathrm{p}}^{\mathrm{cmd}}(t) \tag{20}$$

and the controller must guarantee asymptotic tracking convergence,

$$\lim_{t\to\propto} e_{\mathrm{q}}(t) = 0 \tag{21}$$

while maintaining bounded hydraulic actuation and preserving robustness under severe rheological disturbances.

*5.3. Sliding Surface Design*

To improve robustness and eliminate steady-state tracking deviations under uncertain hydraulic conditions, a first-order integral sliding manifold is adopted [24], [25], [25], [30], [48]. The integral sliding state is represented as

$$\frac{\mathrm{d}\xi_{\mathrm{eq}}(t)}{\mathrm{d}t} = e_{\mathrm{q}}(t) \tag{22}$$

and the sliding manifold is constructed according to

$$s_{\mathrm{q}}(t) = e_{\mathrm{q}}(t) + \lambda_{\mathrm{q}}\cdot\xi_{\mathrm{eq}}(t) \tag{23}$$

where the sliding variable, the integral sliding state, and the sliding-surface gain determine the convergence dynamics of the hydraulic subsystem.

The selected manifold combines proportional and integral error dynamics, thereby improving robustness against slow rheological drift, hydraulic parameter mismatch, and consistency-dependent transport uncertainty. Such integral sliding structures are particularly suitable for slurry-transport systems because they improve low-frequency disturbance rejection while preserving finite-time convergence properties.

*5.4. Equivalent Control Law*

Under ideal sliding conditions, the system trajectory evolves directly on the sliding manifold according to

$$s_q(t) = 0 \tag{24}$$

and

$$\frac{d\xi_{eq}(t)}{dt} = 0 \tag{25}$$

Differentiating the sliding manifold yields

$$\frac{ds_q(t)}{dt} = \frac{de_q(t)}{dt} + \lambda_q \cdot e_q(t) \tag{26}$$

Assuming slowly varying supervisory references,

$$\frac{dq_p^{cmd}(t)}{dt} \approx 0 \tag{27}$$

the tracking-error dynamics become directly coupled with the nonlinear hydraulic transport subsystem. Substituting the nonlinear hydraulic dynamics into the sliding-surface derivative yields the nonlinear control-oriented transport equation governing the SMC synthesis.

The equivalent hydraulic control action is obtained by imposing ideal sliding conditions on the nonlinear transport dynamics, resulting in the nominal hydraulic head required to sustain the desired discharge-flow trajectory under ideal operating conditions,

$$H_{eq}(t) = H_{static}(t) + (C_n(t) + \varepsilon)\cdot(q_p{}^{cmd}(t))^n \tag{28}$$

Equation (28) represents the nominal hydraulic energy necessary to maintain the desired transport condition in the absence of disturbances and uncertainty.

*5.5. Sliding-Mode Switching Law*

To guarantee finite-time convergence toward the sliding manifold, a discontinuous switching component is incorporated into the control structure,

$$H_{sw}(t) = -\,k_{smc}\cdot \mathrm{sat}(s_q(t)/\phi_q(t)) \tag{29}$$

where the switching gain, the boundary-layer thickness, and the saturation function determine the discontinuous correction dynamics.

Accordingly, the complete nonlinear SMC law becomes

$$H_0(t) = H_{eq}(t) - k_{smc}\cdot \mathrm{sat}(s_q(t)/\phi_q(t)) \tag{30}$$

The saturation regularization replaces the ideal sign function to mitigate high-frequency chattering and improve practical implementability in hydraulic actuation systems. This regularization is particularly important in hydraulic transport systems because excessive switching may excite undesirable transient oscillations within the pump and pipeline subsystems.

*5.6. Lyapunov Stability Analysis*

To demonstrate closed-loop stability, the Lyapunov candidate function is selected as

$$V(s_q(t)) = 0.5 \cdot s_q^2(t) \tag{31}$$

whose derivative is expressed as

$$\frac{\mathrm{d}V(t)}{\mathrm{d}t} = s_q(t) \cdot \frac{\mathrm{d}s_q(t)}{\mathrm{d}t} \tag{32}$$

Substituting the SMC dynamics yields

$$\frac{\mathrm{d}s_q(t)}{\mathrm{d}t} = -\frac{k_{smc}}{\tau_p} \cdot \mathrm{sat}\left(\frac{s_q(t)}{\phi_q(t)}\right) + \Delta(t) \tag{33}$$

where $\Delta(t)$ represents the lumped uncertainty associated with:

- consistency reconstruction,
- rheological mismatch,
- hydraulic parameter variation,
- drainability fluctuations,
- channeling disturbances,
- and unmodeled transport effects.

Assuming bounded uncertainty,

$$|\Delta(t)| \leq \Delta_{max} \tag{34}$$

the Lyapunov derivative satisfies

$$\frac{\mathrm{d}V(t)}{\mathrm{d}t} \leq -\frac{k_{smc}}{\tau_p} \cdot |s_q(t)| + |s_q(t)| \cdot \Delta_{max} \tag{35}$$

Therefore,

$$\frac{\mathrm{d}V(t)}{\mathrm{d}t} \leq 0 \text{ if } -k_{smc} > \tau_p \cdot \Delta_{max} \tag{36}$$

Under this condition, the sliding manifold becomes globally attractive and the tracking error converges asymptotically toward the boundary layer,

$$|s_q(t)| \leq \phi_q(t) \tag{37}$$

*5.7. Robustness Against Process Disturbances*

The proposed SMC framework exhibits robustness against multiple nonlinear disturbances inherent to industrial pulp blow-line operation.

The drainability coefficient modifies the entrained-liquor transport dynamics according to

$$f_{liq}(t) = (1 - k_{ch}(t)) \cdot (1 - y_K(t) \cdot C(t)) \cdot \rho_{mix}(t) \cdot (1 - C(t)) \cdot q_p(t) \tag{38}$$

Variations in drainability modify liquid retention, thereby altering slurry consistency and hydraulic transport resistance.

Similarly, the channeling factor introduces preferential hydraulic pathways affecting liquor redistribution and consistency evolution throughout the discharge process. These disturbances generate severe nonlinear hydraulic perturbations that are effectively rejected by the discontinuous sliding action.

Inlet dilution-flow disturbances additionally modify the free-liquor inventory and consequently alter the rheological condition and hydraulic transport capability of the suspension. The proposed SMC structure compensates these disturbances directly through the hydraulic flow-tracking correction mechanism.

### *5.8. Chattering Mitigation*

Classical sliding-mode controllers inherently suffer from high-frequency oscillations associated with discontinuous switching [24], [25], [30], [48]. To reduce chattering effects, the proposed control structure incorporates:

- boundary-layer saturation regularization,
- hydraulic relaxation dynamics,
- actuator first-order filtering,
- reference conditioning,
- and supervisory flow limitation.

The hydraulic actuator dynamics are represented as

$$\tau_{\mathrm{H}} \cdot \frac{\mathrm{d}H_0(t)}{\mathrm{d}t} = H_{0\mathrm{s}}(t) - H_0(t) \tag{39}$$

where the actuator time constant introduces additional filtering into the hydraulic subsystem. This filtering significantly attenuates switching-induced oscillations and improves numerical stability during long-time nonlinear simulations.

### *5.9. Supervisory Consistency Protection*

To avoid unstable operation under excessive consistency conditions, a nonlinear supervisory protection layer is incorporated into the control structure.

The consistency-limiting sigmoid function is defined as

$$\sigma_{\mathrm{C}}(t) = \frac{1}{1 + \mathrm{e}^{-\alpha_{\mathrm{C}} \cdot (C_{\max} - C(t))}} \tag{40}$$

while the protected discharge reference becomes

$$q_{\mathrm{P}}^{*}(t) = \sigma_{\mathrm{C}}(t) \cdot q_{\mathrm{P}}^{\mathrm{ref}}(t) \tag{41}$$

Consequently, when the suspension approaches unsafe high-consistency operating conditions, the discharge reference is automatically reduced, thereby preventing hydraulic overload and excessive rheological resistance within the blow-line subsystem.

## **6.0. Simulation Results**

The proposed nonlinear batch digester discharge model and the associated SMC architecture were evaluated through transient simulations performed in MATLAB–Simulink under nonlinear rheological transport conditions, consistency-dependent hydraulic resistance, drainability disturbances, channeling effects, and inlet-flow perturbations. The simulations were specifically designed to assess the dynamic robustness, hydraulic stability, and discharge-flow tracking capability of the proposed SMC framework during long-time blowdown operation under highly coupled multiphase transport conditions.

The nonlinear process model was implemented using the robust hydraulic regularization and protection mechanisms described previously, including hydraulic-head protection, bounded hydraulic resistance, consistency saturation, density limitation, bounded internal transport flows, and actuator filtering. Due to the strong nonlinear coupling and the differential-algebraic structure of the process, the stiff solver ODE15s was employed to ensure numerical convergence and stability throughout the transient simulations. The principal simulation parameters employed during the transient analysis are summarized in Table 1.

The imposed disturbance scenario incorporated three principal transient perturbations representative of industrial batch digester operation. First, the channeling factor was modified from ($k_{ch}(t) = 0.50$) to ($k_{ch}(t) = 0.80$) at ($t = 2.0 \cdot 10^4$ s}) in order to simulate the formation of preferential hydraulic pathways through the fiber matrix. Subsequently, the drainability coefficient was increased from ($y_K(t) = 0.20$) to ($y_K(t) = 0.50$) at ($t = 5.0 \cdot 10^4$ s}) to represent severe permeability degradation and increased liquor retention inside the pulp network. Finally, the inlet dilution-liquor flow was increased from ($f_{in}(t) = 1.0 \cdot 10^{-4}$, m$^3$/s) to ($f_{in}(t) = 1.5 \cdot 10^{-4}$, m$^3$/s) at ($t = 6.0 \cdot 10^4$, s) in order to evaluate the robustness of the hydraulic controller under varying dilution conditions.

Fig. 3(a) presents the imposed disturbance profiles associated with the channeling factor ($k_{ch}(t)$) and the drainability coefficient ($y_K(t)$). The introduced perturbations generate substantial nonlinear variations in liquor redistribution, suspension permeability, and effective hydraulic resistance. Despite the severe rheological transitions produced by these disturbances, the proposed SMC framework preserves stable hydraulic operation throughout the complete blowdown sequence.

Fig. 3(b) illustrates the principal discharge-flow tracking dynamics, including the commanded discharge flow ($q_p^{cmd}(t)$), the algebraic hydraulic prediction ($q_p^{alg}(t)$), and the measured discharge flow ($q_p(t)$). The results demonstrate excellent tracking capability with negligible steady-state deviation and strong transient robustness despite the nonlinear hydraulic disturbances and consistency-dependent rheological variations. The response confirms that the proposed SMC strategy successfully compensates the nonlinear pressure–flow coupling introduced by the evolving hydraulic resistance of the pulp suspension.

The transient evolution of the sliding manifold is shown in Fig. 3(c). The response demonstrates rapid convergence of the sliding surface toward a narrow bounded region around zero, thereby confirming that the closed-loop trajectory successfully reaches and remains within the designed sliding regime. This behavior validates the robustness and finite-time convergence characteristics associated with the proposed sliding-mode formulation.

Similarly, Fig. 3(d) presents the discharge-flow tracking error ($e_q(t)$). The results indicate that the tracking error remains bounded and rapidly attenuated following each disturbance event, demonstrating the capability of the proposed controller to reject severe nonlinear perturbations associated with channeling, drainability degradation, and inlet-flow variations. Furthermore, the observed error dynamics confirm that the selected boundary-layer regularization effectively mitigates high-frequency oscillations without compromising tracking performance.

Table 1. Simulation Process Variable

| Parameter | Description | Value |
|---|---|---|
| $\rho_s$ | Dry-fiber density | 1050 [kg/m$^3$] |
| $\rho_{fl}$ | Free-liquor density | 1100 [kg/m$^3$] |
| $n$ | Herschel–Bulkley flow index | 0.75 |
| $K_{ref}$ | Reference hydraulic resistance coefficient | $8 \cdot 10^3$ |
| $C_{ref}$ | Reference consistency | 0.10 |
| $\alpha_C$ | Consistency-resistance exponent | 2.0 |
| $\tau_y$ | Yield stress | 50 [Pa] |
| $K_{HB}$ | Herschel–Bulkley consistency index | 75 [Pa/s$^n$] |
| $D_{pipe}$ | Blow-line diameter | 0.20 [m] |
| $L_{eff}$ | Effective pipeline length | 20 [m] |
| $H_0^{max}$ | Maximum hydraulic head | 120 [m] |
| $\tau_H$ | Pump actuator time constant | 300 [s] |
| $q_p^{max}$ | Maximum discharge flow | 0.004 [m$^3$/s] |
| $M_s(0)$ | Initial dry-fiber mass | 2500 [kg] |
| $M_{fl}(0)$ | Initial free-liquor mass | 25000 [kg] |
| $\lambda_q$ | Sliding-manifold gain | $1 \cdot 10^{-4}$ |
| $k_{smc}$ | SMC switching gain | 3 |
| $\phi_q$ | Boundary-layer thickness | $5 \cdot 10^{-4}$ |

The transient evolution of the dry-fiber and free-liquor inventories is illustrated in Fig. 3(e), where the dynamics of ($M_s(t)$) and ($M_{fl}(t)$) are presented. Since the proposed SMC strategy regulates exclusively the hydraulic discharge flow, these internal inventories evolve naturally according to the intrinsic nonlinear process dynamics. The observed inventory depletion reflects the coupled interaction between slurry transport, liquor redistribution, and rheological evolution occurring during the blowdown operation.

Fig. 3(f) shows the transient evolution of the pulp consistency ($C(t)$). As expected, the consistency progressively increases during discharge due to the depletion of free liquor and the resulting increase in solid concentration inside the digester. This behavior is particularly relevant because the consistency directly governs the rheological transport characteristics and therefore strongly influences the effective hydraulic resistance of the blow-line subsystem.

The corresponding consistency-dependent hydraulic resistance ($C_n(t)$) is presented in Fig. 3(g). The results confirm the strong nonlinear coupling between slurry composition and hydraulic transport resistance. As the consistency increases, the effective hydraulic resistance grows significantly, thereby increasing the hydraulic energy required to sustain the discharge process. This behavior is fully consistent with the rheological characteristics of concentrated pulp suspensions reported in the literature.

Fig. 3(h) presents the transient dynamics of the internal transport flows, including the dry-fiber transport flow ($f_s(t)$), the entrained-liquor transport flow ($f_{liq}(t)$), the extracted free-liquor flow ($f_{fl}(t)$), and the inlet dilution-liquor flow ($f_{in}(t)$). The results demonstrate the strong interaction among the internal transport mechanisms, particularly during the imposed disturbance intervals. Variations in drainability and channeling directly affect the liquor redistribution dynamics, thereby modifying the internal transport conditions of the suspension.

The transient evolution of the static hydraulic head ($H_{static}(t)$) is illustrated in Fig. 3(i). The response remains smooth and bounded throughout the complete simulation horizon, thereby confirming the stability of the hydraulic reconstruction and the effectiveness of the implemented numerical protection mechanisms. Since the static hydraulic head depends directly on the evolving mixture density, this variable constitutes an important internal disturbance acting on the nonlinear hydraulic subsystem.

Fig. 3(j) presents the transient evolution of the effective slurry density ($\rho_{mix}(t)$). The obtained response demonstrates stable phase reconstruction and confirms that the nonlinear mixture-density formulation remains numerically robust during severe rheological transitions and long-time discharge simulations.

Finally, Fig. 3(k) illustrates the hydraulic control actions generated by the proposed SMC architecture, including the applied hydraulic head ($H_0(t)$), the equivalent hydraulic control action ($H_{eq}(t)$), and the saturated control signal ($H_{0s}(t)$). The results demonstrate smooth hydraulic actuation without evidence of severe chattering or unstable switching behavior. The implemented boundary-layer saturation and actuator dynamics successfully attenuate high-frequency oscillations while preserving robust discharge-flow tracking performance.

Overall, the transient simulation results presented in Fig. 3(a)–(k) confirm that the proposed nonlinear SMC framework provides robust and numerically stable discharge-flow regulation under severe rheological uncertainty, nonlinear hydraulic coupling, and time-varying transport disturbances. The controller maintains excellent tracking performance, preserves bounded hydraulic actuation, and guarantees stable blow-line operation despite the highly nonlinear and strongly coupled behavior characteristic of industrial batch digester discharge systems.

Additionally, the energetic performance associated with the proposed SMC-regulated discharge operation is presented in Fig. 4. The results provide further insight into the coupled interaction between hydraulic actuation, rheological dissipation, transport efficiency, and cumulative energetic consumption during the nonlinear blowdown process. The energetic analysis confirms that the discharge dynamics are dominated by a highly transient hydraulic regime in which the principal energetic demand occurs during the early stages of slurry mobilization and transport acceleration.

Fig. 4(a) shows the instantaneous hydraulic efficiency ($\eta_h(t)$). The response demonstrates that the transport efficiency decreases progressively during the discharge operation as the internal slurry consistency increases and the available free liquor is depleted. This behavior is physically consistent with the increasing rheological resistance and reduced suspension mobility characteristic of concentrated pulp transport systems. As the slurry becomes progressively more compact, larger hydraulic energy is required to sustain equivalent transport conditions, thereby reducing the effective energetic efficiency of the hydraulic subsystem.

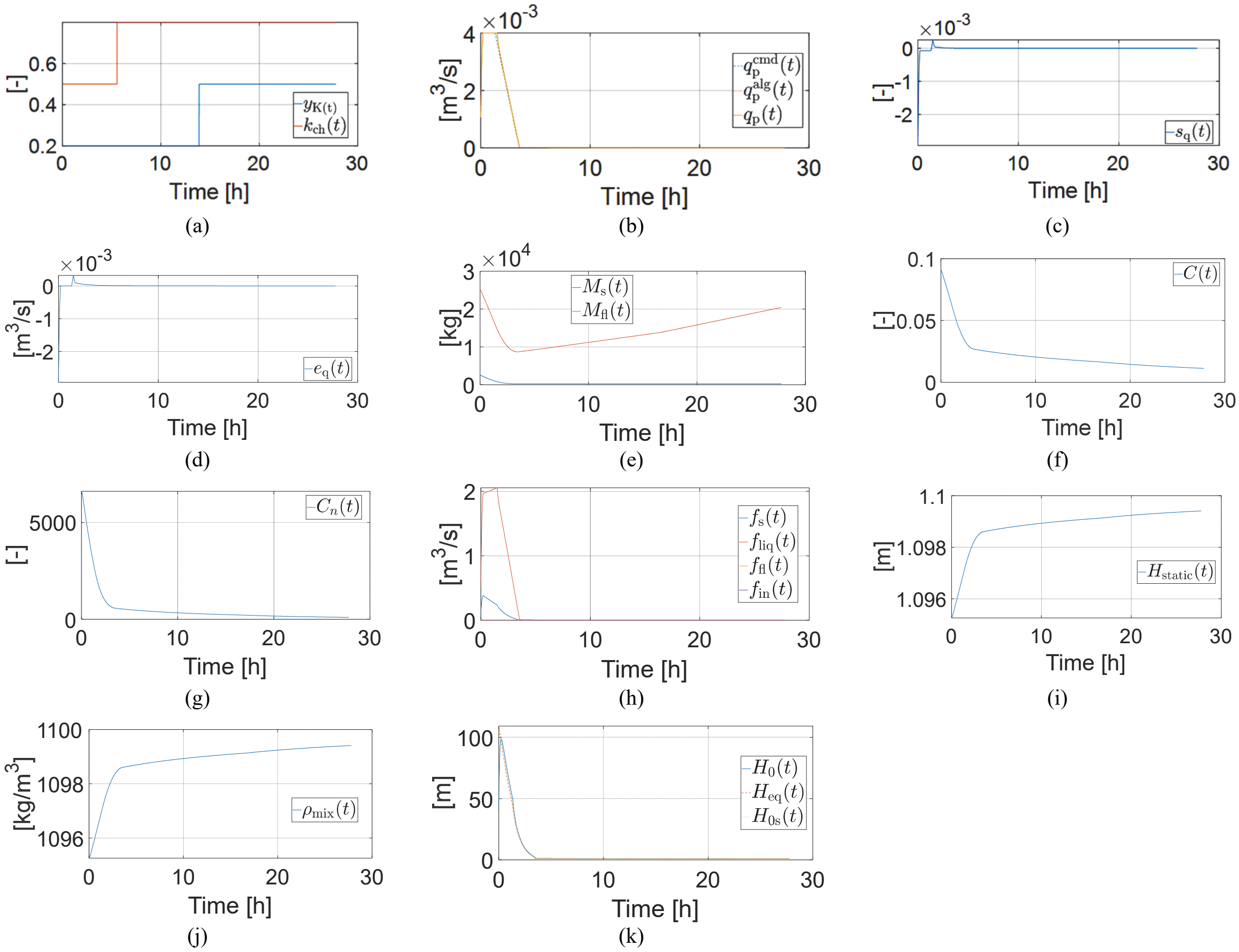

Fig. 3. Transient simulation results of the proposed nonlinear batch digester discharge system under the SMC strategy. (a) Imposed disturbances $k_{ch}(t)$ and $y_K(t)$. (b) Discharge-flow tracking performance. (c) Sliding manifold dynamics $s_q(t)$. (d) Tracking error evolution $e_q(t)$. (e) Dry-fiber and free-liquor mass dynamics. (f) Consistency evolution $C(t)$. (g) Consistency-dependent hydraulic resistance $C_n(t)$. (h) Transport-flow dynamics. (i) Static hydraulic head $H_{static}(t)$. (j) Effective slurry density $\rho_{mix}(t)$. (k) Hydraulic control actions generated by the proposed SMC strategy.

The corresponding hydraulic power consumption is illustrated in Fig. 4(b). A pronounced transient power peak appears during the initial discharge stage, indicating that the highest energetic demand occurs during the rapid acceleration and mobilization of the concentrated pulp suspension inside the blow-line subsystem. Subsequently, the hydraulic power decreases rapidly as the discharge flow attenuates and the transport process approaches low-flow operating conditions.

Fig. 4(c) compares the useful transport power and the estimated electrical power demand. The results demonstrate that the electrical power remains consistently larger than the useful transport contribution due to hydraulic conversion losses, rheological dissipation, and pump–motor efficiency limitations. Nevertheless, the proposed SMC architecture preserves stable energetic behavior throughout the complete discharge sequence without generating excessive hydraulic oscillations or unstable energetic transients.

The rheological dissipation functional ($\Phi_V(t)$) is presented in Fig. 4(d). The results indicate that the dominant viscous dissipation occurs during the initial transient interval where the slurry experiences the highest shear deformation rates and strongest hydraulic acceleration. As the discharge flow decreases, the shear rate and internal rheological deformation diminish substantially, thereby reducing the associated dissipation losses.

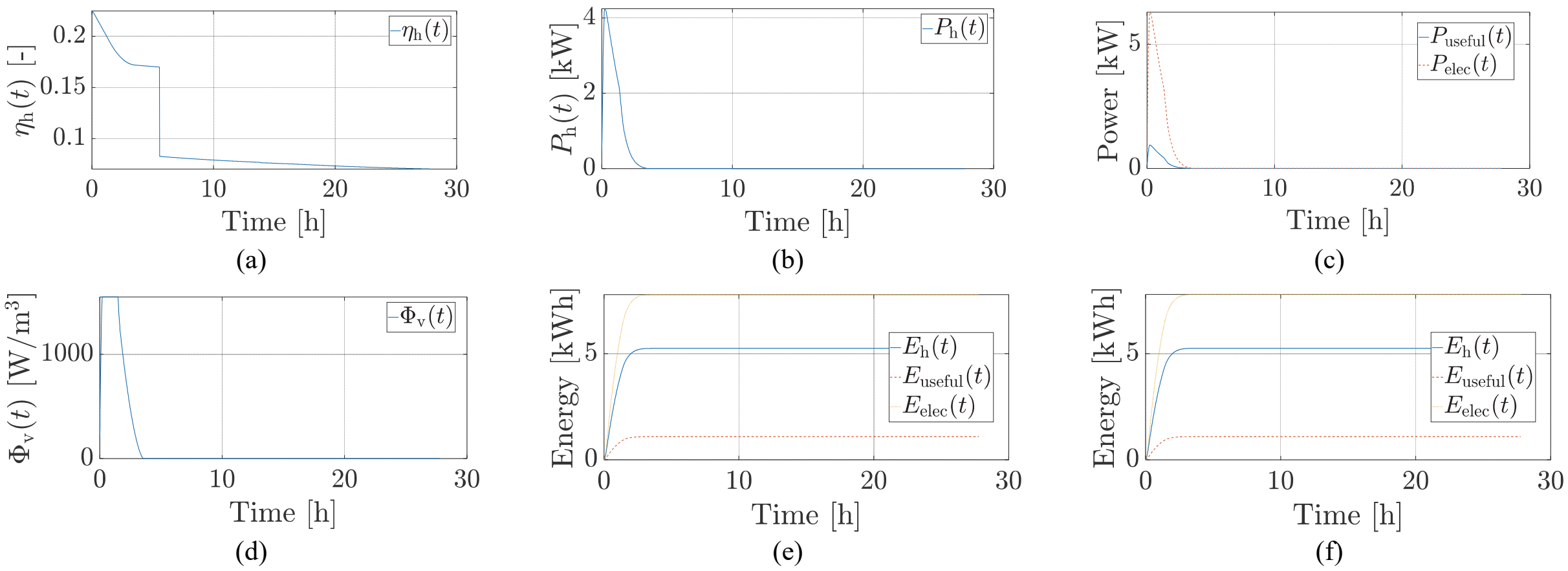

Fig. 4. Thermodynamic and hydraulic efficiency analysis of the proposed nonlinear batch digester discharge system under SMC regulation. (a) Instantaneous hydraulic efficiency $\eta_h(t)$. (b) Hydraulic power consumption $P_h(t)$. (c) Useful and electrical power evolution $P_{useful}(t)$ and $P_{elec}(t)$. (d) Rheological dissipation functional $\Phi_v(t)$. (e) Cumulative hydraulic and useful energy consumption $E_h(t)$ and $E_{useful}(t)$. (f) Cumulative electrical energy demand $E_{elec}(t)$.

The cumulative energetic responses are illustrated in Fig. 4(e) and Fig. 4(f). The hydraulic, useful, and electrical energy curves exhibit rapid accumulation during the active blowdown interval followed by asymptotic stabilization as the transport process approaches near-zero flow conditions. These results confirm that the majority of the energetic consumption is concentrated within the initial transient discharge period, which is fully consistent with the strongly nonlinear hydraulic transport behavior governing concentrated pulp-slurry systems.

Overall, the energetic analysis presented in Fig. 4 confirms that the proposed nonlinear SMC framework maintains stable hydraulic transport performance while preserving bounded energetic consumption under severe rheological nonlinearities, evolving consistency conditions, channeling disturbances, and dynamic drainability variations. The obtained results additionally demonstrate that the proposed controller successfully mitigates excessive hydraulic energy demand and preserves robust energetic behavior throughout the complete nonlinear blowdown operation.

Additionally, the geometric interpretation of the proposed sliding-mode manifold is illustrated in Fig. 5, where the three-dimensional sliding surface associated with the discharge-flow controller is presented as a function of the tracking error $e_q(t)$ and the integral sliding-state variable $\xi_{eq}(t)$. The surface representation provides important insight into the nonlinear convergence mechanism governing the proposed SMC strategy and the associated closed-loop stability properties. The generated manifold defines the admissible dynamic trajectories of the controlled hydraulic system and establishes the robust attraction region toward the desired sliding regime. The zero-manifold condition $s_q(t) = 0$ corresponds to the ideal operating trajectory in which the discharge-flow tracking dynamics become asymptotically constrained to the designed sliding surface despite the presence of nonlinear rheological uncertainties and hydraulic disturbances.

As observed in Fig. 5, the proposed sliding surface exhibits a well-defined linear hyperplane structure with smooth geometric continuity over the complete error domain. This behavior confirms that the selected manifold formulation generates a globally continuous switching geometry suitable for robust hydraulic regulation under strongly nonlinear operating conditions. The inclination of the manifold is primarily governed by the selected integral sliding gain $\lambda_q$, which determines the relative weighting between the instantaneous flow-tracking error and the accumulated integral dynamics. Consequently, the manifold simultaneously incorporates transient and long-term tracking information, thereby improving disturbance rejection capability and eliminating steady-state deviation during severe nonlinear blowdown operation.

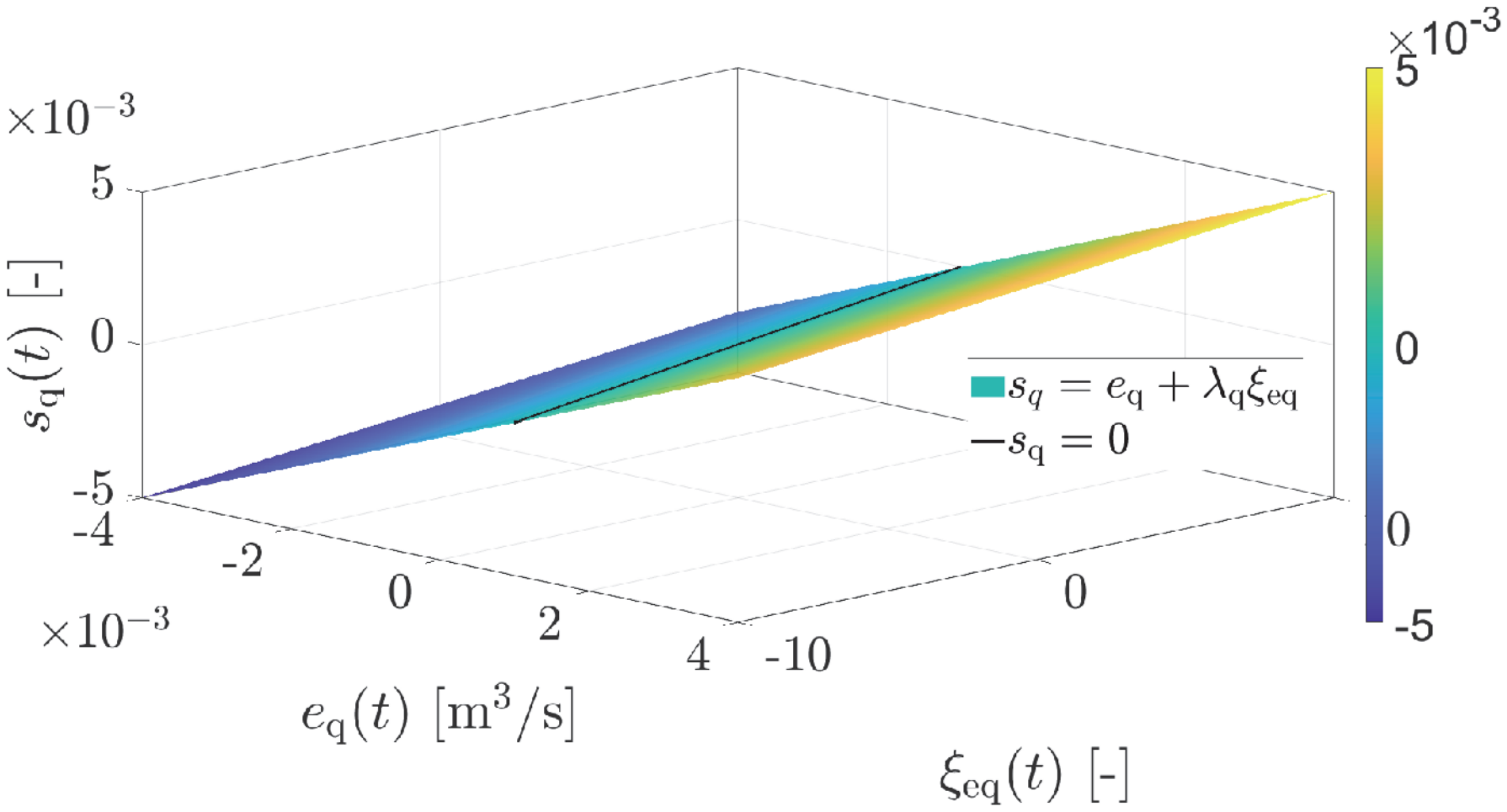

Fig. 5. Three-dimensional sliding manifold associated with the proposed SMC strategy, showing the relationship between the discharge-flow tracking error $e_q(t)$, the integral sliding state $\xi_{eq}(t)$, and the sliding surface $s_q(t)$. The zero-contour represents the ideal sliding regime $s_q(t) = 0$ used for robust nonlinear discharge-flow regulation.

The black contour corresponding to the condition $s_q(t) = 0$ represents the ideal sliding trajectory enforced by the proposed SMC law. Once the closed-loop hydraulic dynamics reach this region, the system behavior becomes dominated by the equivalent reduced-order dynamics associated with the sliding regime. This characteristic constitutes one of the principal advantages of SMC since the closed-loop system becomes substantially less sensitive to parametric uncertainty, rheological variability, and external transport disturbances after reaching the manifold. Therefore, the surface shown in Fig. 5 directly illustrates the robustness mechanism responsible for the stable discharge-flow tracking performance previously observed in Fig. 3(b)–(d).

Furthermore, the geometric smoothness observed around the sliding region confirms the effectiveness of the implemented boundary-layer regularization used to mitigate classical chattering phenomena. Instead of generating discontinuous switching trajectories, the proposed saturation-based implementation produces continuous manifold convergence while preserving robust hydraulic regulation. This behavior is particularly important in industrial hydraulic transport systems because excessive switching activity may induce undesirable pump oscillations, actuator stress, and unstable pressure transients inside the blow-line subsystem.

Overall, the sliding-surface representation presented in Fig. 5 provides additional validation of the theoretical robustness and convergence characteristics of the proposed nonlinear SMC architecture. The obtained manifold geometry confirms that the controller generates a stable and well-conditioned attraction region capable of preserving bounded discharge-flow regulation under the highly nonlinear rheological and hydraulic conditions characteristic of industrial batch digester blowdown operations.

## 7.0. Conclusions

This work presented a comprehensive nonlinear multiphysics modeling and robust SMC framework for the dynamic analysis and hydraulic regulation of batch digester discharge systems operating under highly nonlinear rheological transport conditions. The proposed formulation integrated coupled mass balances, consistency-dependent hydraulic resistance, nonlinear hydrotransport dynamics, drainability behavior, channeling effects, and non-Newtonian slurry rheology into a unified nonlinear differential-algebraic representation capable of reproducing the dominant physical interactions governing industrial blowdown operations.

The developed model demonstrated that the discharge dynamics of batch digesters are intrinsically governed by a strong nonlinear coupling between slurry consistency evolution, hydraulic transport capability, liquor redistribution, and rheological resistance. In particular, the results confirmed that relatively small variations in consistency and permeability conditions may produce substantial modifications in the effective hydraulic resistance, discharge-flow capability, and energetic behavior of the transport subsystem. The inclusion of phenomenological channeling and drainability mechanisms additionally allowed the proposed framework to reproduce important industrial transport phenomena commonly neglected in simplified pulp-slurry discharge formulations.

From a control-system perspective, the proposed SMC architecture demonstrated excellent robustness against severe nonlinear disturbances, rheological uncertainty, hydraulic parameter variation, and evolving transport conditions characteristic of industrial blow-line operation. The transient simulation results confirmed that the controller successfully maintained stable discharge-flow regulation despite the presence of strong nonlinear hydraulic coupling, dynamic consistency variation, inlet-flow perturbations, channeling disturbances, and drainability degradation. Furthermore, the implemented boundary-layer regularization and hydraulic actuator dynamics effectively mitigated chattering phenomena while preserving robust sliding-mode convergence and bounded hydraulic actuation.

The obtained transient responses additionally demonstrated that the proposed nonlinear framework remained numerically stable during long-time simulations despite the severe stiffness and nonlinear differential-algebraic characteristics associated with concentrated pulp-suspension transport. The implemented numerical regularization strategies, hydraulic protection mechanisms, bounded transport reconstruction, and actuator filtering significantly improved simulation robustness and prevented the occurrence of singularities, hydraulic runaway, and non-finite dynamic states during extreme operating conditions. Consequently, the proposed methodology establishes a stable computational platform suitable for future large-scale simulation and advanced control studies involving nonlinear pulp-processing systems.

The energetic analysis further demonstrated that the discharge process operates under a highly transient hydraulic regime dominated by strong rheology-dependent dissipation and consistency-driven transport resistance.

The results confirmed that the majority of the energetic demand occurs during the initial slurry acceleration and mobilization stages, whereas the progressive increase in pulp consistency substantially reduces transport efficiency due to elevated hydraulic resistance and non-Newtonian dissipation effects. Nevertheless, the proposed SMC structure preserved bounded energetic behavior and avoided excessive hydraulic oscillations throughout the complete blowdown operation.

The geometric analysis of the sliding manifold additionally validated the theoretical robustness properties of the proposed SMC strategy. The generated sliding surface established a well-conditioned attraction region capable of maintaining stable closed-loop hydraulic regulation under severe nonlinear disturbances and uncertain rheological operating conditions. The obtained manifold structure confirmed the capability of the proposed controller to preserve finite-time convergence and disturbance rejection despite the highly nonlinear behavior characteristic of industrial pulp-slurry hydrotransport systems.

It is important to emphasize that the present investigation constitutes a proof-of-concept study focused on the theoretical development and numerical evaluation of a nonlinear modeling and robust control framework for batch digester discharge operations. Therefore, the primary contribution of this work lies in the formulation, simulation, and control-oriented analysis of the proposed nonlinear transport model rather than in the experimental validation of a specific industrial installation. Nevertheless, the obtained results demonstrate that the proposed methodology provides a physically consistent and computationally robust foundation for future industrial implementation and experimental investigation.

Future research should therefore focus on industrial parameter identification, pilot-scale experimental validation, distributed transport modeling, multiphase CFD-assisted reconstruction, advanced observer design, and the incorporation of predictive and adaptive nonlinear control methodologies. Additionally, future developments may incorporate thermal transport effects, delignification kinetics, blow-line pressure-wave propagation, and distributed fiber-network deformation mechanisms in order to achieve a more comprehensive digital representation of industrial batch digester discharge systems.